# Natural Time, Nowcasting and the Physics of Earthquakes: Estimation of Seismic Risk to Global Megacities

John B Rundle[1,2,3,4], Molly Luginbuhl[1], Alexis Giguere[1], Donald L. Turcotte[3]

[1] Department of Physics
University of California, Davis, CA
[2] Santa Fe Institute
Santa Fe, NM
[3] Department of Earth and Planetary Science
University of California, Davis, CA
[4] Open Hazards Group
Davis, CA

## Abstract

*Natural Time* ("NT") refers to the concept of using small earthquake counts, for example of M>3 events, to mark the intervals between large earthquakes, for example M>6 events. The term was first used by (Varotsos et al., 2005) and later by (Holliday et al., 2006) in their studies of earthquakes. As we discuss in this paper, it is particularly useful in describing complex stochastic nonlinear systems characterized by fat-tail statistics rather than Gaussian normal statistics. In this paper we discuss ideas and applications arising from the use of NT to understand earthquake dynamics. The usual end-user applications of fault-based studies are often applied to risk of a particular geographic location, so it seems best to start the analysis with that geographic region. Rather than focus on an individual earthquake *faults*, we have found it more productive to focus on a defined *local geographic region* surrounding a particular location. This local region is considered to be embedded in a larger regional setting from which we accumulate the relevant statistics. From this different philosophical point of view, we first discuss methods to use NT, counts of small earthquakes, to evaluate the current state of a regional collection of faults. We then use these concepts to first discuss the nucleation physics of large earthquakes. We introduce the idea of *nowcasting*, a term originating from economics and finance. The goal of nowcasting is to determine the current state of the fault system, or put another way, the current state of progress through the earthquake cycle. This is in contrast to *forecasting*, which is the calculation of probabilities of future large earthquakes. Finally, we apply the nowcasting idea to the practical development of methods to estimate the current state of risk for dozens of the world's seismically exposed megacities, defined as cities having populations of over 1 million persons. We compute a ranking of these cities based on their current nowcast value, and discuss the advantages and limitations of this approach. We note explicitly that the nowcast method is *not a model*, in that there are no free parameters to be fit to data. Rather, the method is simply a presentation of statistical data, which the user can interpret.

# Introduction

Natural time is a term first used by (Varotsos et al., 2005, 2011)and subsequently by (Holliday et al., 2006). It builds on the idea that driven threshold systems such as earthquake fault systems often display a power-law distribution of event sizes or magnitudes. While these bursts of activity are observed at all scales, the largest events are usually of most interest. For earthquakes, these largest events are the magnitude 6+ events that cause the most damage and injuries. Interspersed between these largest events are many smaller events of varying sizes and magnitudes.

Taken together, these small and large events are distributed in a scale-invariant power-law statistical distribution of magnitude. The Gutenberg-Richter magnitude-frequency law (Gutenberg and Richter, 1942; Scholz, 1990) is a simple model of this distribution which is found to be applicable over large spatial domains and over long time intervals. The GR model has two parameters, a and b, which must be fit to the observed data:

$$Log_{10} N = a - bM \quad (1)$$

Here $N$ is the number or frequency of earthquakes having magnitudes larger than $M$. Typically, $b \sim 1$.

Over smaller spatial domains and shorter time intervals, the actual statistics of the observed number or frequency of earthquakes can depart considerably from the simple model (1). A good example is shown in Figure 1a,b . On the left in Figure 1a we see a map of a large region surrounding the city of city of San Diego, USA, between latitudes 24º N and 43º N Latitude, and between 128º W and 110º W Longitude. In the center of the map is a circle of radius 400 km surrounding the city of San Diego. We then construct the Gutenberg-Richter (GR) number-magnitude statistics in Figure 1b. The statistics represented by the square symbols are derived from all earthquakes in the large region having magnitudes greater than the completeness magnitude of M3.0 since 1970, at depths less than 200 km, contained within the ANSS Composite Catalog (ANSS, 2017).

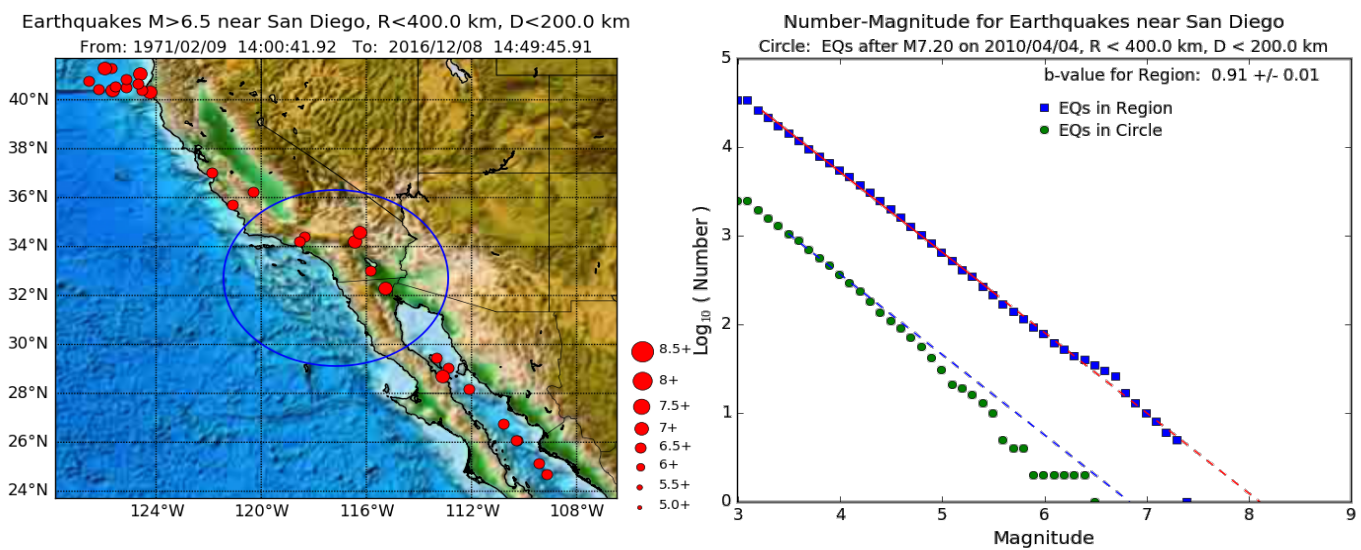

Figure 1. a) Map of earthquakes having magnitude $M \geq 6.5$ near San Diego since 1970. Circle centered on San Diego has radius $R = 400$ km. b) GR number-magnitude statistics. The upper blue square symbols are all earthquakes $M \geq 3$ for the region as a whole since 1970. The lower green circles are all earthquakes $3 \leq M < 6.5$ since the last $M \geq 6.5$ earthquake, which was the M7.2 El Major-Cucapah earthquake on 4/4/2010.

The red line in Figure 1b is the best fitting GR model, where the b-value (slope) as shown is $0.91 \pm 0.01$. The fit of the model is taken between magnitudes M3.25 and M5.5, and we show the extension of the fit line to the X-axis, intersecting that axis at M8.1. The round symbols record similar data for all small earthquakes occurring only after the most recent $M \geq 7.0$ earthquake within the 400 km radius circle surrounding San Diego. The blue dashed line is a GR line having the same slope as the red line for the large region $b$=0.91.

It can be seen that, whereas earthquake symbols in the large region closely follow the red dashed GR line up to M7.5, recent events in the circular region closely follow the blue dashed line for events less than M4.5. For earthquakes having M>4.5, the round symbols fall below the blue dashed line. Thus it would appear that there is a deficit of larger earthquakes that will be eventually filled in by the occurrence of large events, assuming that the GR statistics are the same for both the large and small regions.

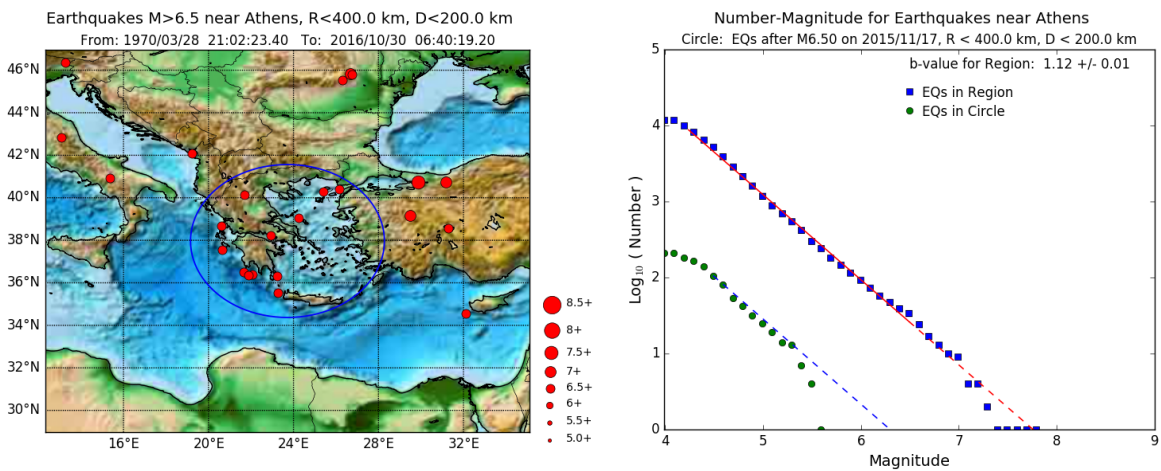

Figure 2. a) Map of earthquakes having magnitude $M \geq 6.5$ near Athens, Greece since 1970. Circle centered on Athens has radius $R = 400$ km. b) GR number-magnitude statistics. The upper blue square symbols are all earthquakes $M \geq 4$ for the region as a whole since 1970. The lower green circles are all earthquakes $4 \leq M < 6.5$ since the last $M \geq 6.5$ earthquake, an M6.5 event on 11/17/2015.

There are many other similar examples. In figure 2a,b we show images for Athens, Greece. Here we assume a completeness magnitude of M4.0, and show both the large region, together with the 400 km radius circular region surrounding the city of Athens. A similar pattern of deficit of larger earthquakes (M>5.5) can be seen for the local events within the circle. It can also be seen that a better estimate of the completeness magnitude is about M~4.5. Yet another example was shown in Figure 1 of Holliday et al. (2016) for the Tokyo, Japan region.

## Statistical Physics of Seismicity: Ergodic Property and Phase Transitions

The fundamental assumption of Holliday et al. (2016) is that a *deficiency* of large earthquakes within a local region contained within a seismically active larger region will eventually be filled in by the occurrence of large earthquakes. The idea is that the statistics of smaller regions over long times will be the same as the statistics of the larger region over large spatial domains and long times. The basis is the idea that seismic activity is characterized by *ergodic dynamics* (Ferguson et al., 1999; Tiampo et al.,. 2003, 2007). Ergodic behavior has been demonstrated for observed seismicity in locations as diverse as California, the Iberian peninsula, and Eastern Canada.

The ergodic property of seismicity was established using the metric published by Thirumalai and Mountain (1989, 1992). An ergodic system is one in which ensemble averages yield the same result as time averages. In our case, the ensemble average is equivalent to a spatial average, inasmuch as a large seismically active region is regarded as being made up of non-overlapping subdomains of smaller seismically active regions.

The physical picture, common in statistical mechanics of complex systems is discussed in the papers by (Ferguson et al., 1999; Tiampo et al.,. 2003, 2007). The system evolves on a time-dependent energy *landscape* in which the system resides in one of a large number of scale-invariant free energy minima or "wells" that dot the landscape. Fluctuations within the energy well are associated with the scale invariant, power-law, small magnitude end of the magnitude-frequency distributions. As the topography of the landscape is constantly evolving, there will occasionally be times when the system escapes from its local energy well in a nucleation event, subsequently landing in another nearby energy well. The nucleation event is associated with a large magnitude earthquake.

This picture of a nucleation event is entirely general, and applies to the dynamics of many statistical physics systems characterized by first order phase transitions (Rundle et al., 2003). First order transitions are the mechanisms by which complex dynamical threshold systems evolve by means of sudden, spontaneous events. We note that one of the signatures of fluctuations in energy landscapes is the characteristic, statistical distribution of seismicity averaged over many cycles of activity. Typically, these fluctuations take the form of the Fisher-Stauffer droplet model, which is common in the study of percolation and phase transitions:

$$n(s) = \frac{n_o}{s^\tau} \exp\left\{-\left(\frac{s}{s_o}\right)^\sigma\right\} \qquad (2)$$

Here, $s$ is the droplet or cluster size, $n_o$ is a constant, $s_o$ is a scale factor, and $\tau$ and $\sigma$ are scaling exponents (Stauffer and Aharony, 1994; Hoshen and Koppelman, ). In addition, $\sigma$ is called the "surface exponent", since its value characterizes properties of the surface of the nucleating droplets in relation to the bulk or interior of the droplet. For the probability density functions characterizing nearest-neighbor site percolation, $s_o$ is related to the value of the occupation probability and thus to the correlation length of the system.

An important special case of the scaling exponents in (1) occurs when the system is mean field, or even near the mean field condition, in which all components of the system interact over long spatial distances (Stauffer and Aharony, 1994; Rundle and Klein, 1993; Rundle et al., 2003; Klein et al., 2000). In that case, when interactions are long range, it is found that $\tau = 2.5, \sigma = 1$. Near mean field systems in particular are interesting because they imply that phenomena such as metastability and hysteresis are to be expected, together with nucleation droplets that have spatial dimensions on the scale of the correlation length (Klein et al., 2007).

## Average Earthquake Statistics

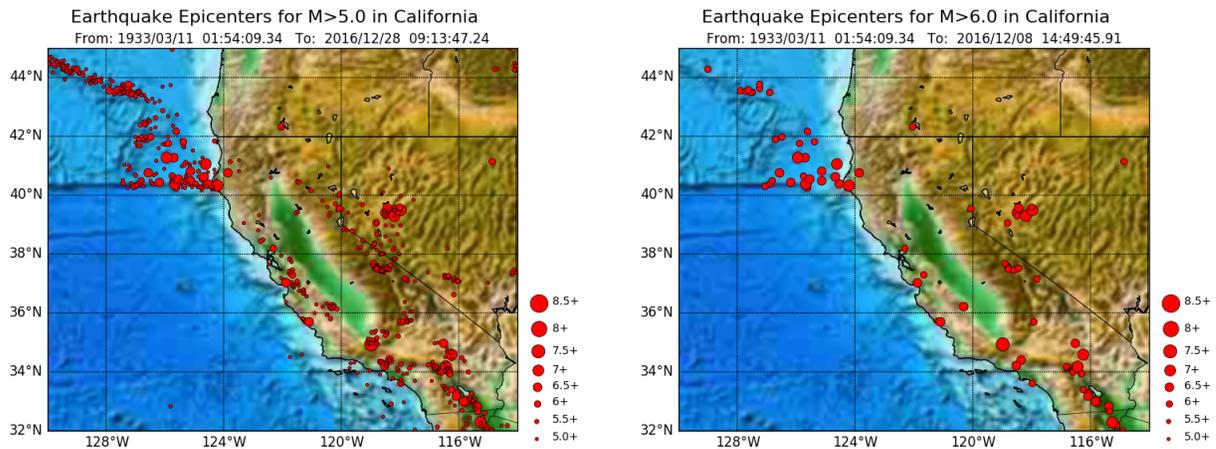

Figure 3. Epicenter maps for California. a) All 732 earthquakes since 1932 for $M \geq 5$. b) All 81 earthquakes since 1932 for $M \geq 6$.

As an example, consider the statistics of seismicity averaged over many earthquake cycles in California, where the data are best (Figure 3). In particular, we will determine the configuration of the *average* earthquake statistics for "small" magnitude events preceding "larger" magnitude events. We assume a completeness magnitude threshold of M3.0, and determine the average earthquake statistics leading up to large events of M≥5.0, and for M≥6.0. Thus we are interested in examining earthquake cycles for earthquakes of M≥5.0 and M≥6.0.

In Figure 3, we show the domain of interest, which consists of the region of California-Nevada-Mexico between 32°N to 45°N Latitude, and between 114°W to 130°W Longitude. Figure 3a shows the 732 M≥5 earthquakes in this region in the catalog, while Figure 3b shows the 81 earthquakes with M≥6. For the M≥5 earthquake cycles, we construct the GR number-magnitude statistics for all the small events $3.0 \leq M < 5.0$, then normalize by divsion by 731, the number of M≥5 earthquake cycles. The results are shown in figure 4 for $\text{Log}_{10}$(number) vs. Magnitude.

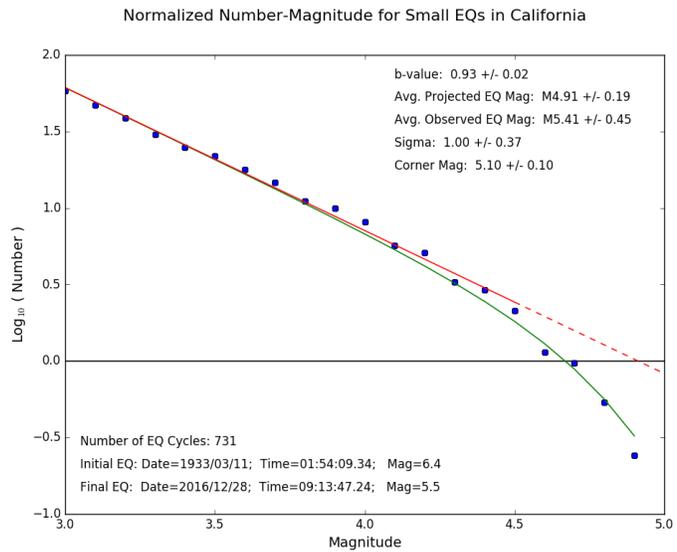

Figure 4. Average GR number-magnitude statistics for small earthquakes $3 \leq M < 5$ in California in the region shown in Figure 3a. Red line shows a fit to the data between magnitudes $3 \leq M \leq 4.5$. Green line is a fit to equation (5), the droplet model.

The red line is a GR fit to the data between the assumed completeness magnitude of M3.0, and M4.5. The green line is a fit to a Fisher-Stauffer droplet-type model equation similar to (2) that we now discuss.

We assume that droplet size can be measured by the elastic energy released during the earthquake, which is the seismic moment of the earthquake. Using base 10 instead of base *e* we have:

$$n(W) = \frac{n_o}{W^{\tau-1}} 10^{\left\{-\left(\frac{W}{W_o}\right)^\sigma\right\}} \tag{3}$$

for the number of earthquakes having seismic moment larger than $W$. Equation (2), which is a survivor function, is then approximately valid in the limit of large $W_o$, which applies to the scaling regime for nucleation and critical phenomena (Stauffer and Aharony, 1994)

Using the definition of seismic moment in terms of magnitude (Scholz, 1990):

$$1.5M = Log_{10}(W) - 9.0 \tag{4}$$

we find:

$$Log_{10}[n(M)] = a - bM - 10^{1.5\sigma(M-M_o)} \tag{5}$$

where:

$$a = Log_{10}(n_o) - 9.0/(\tau - 1) \tag{6}$$

and:

$$b = 1.5/(\tau - 1) \tag{7}$$

We fit equation (3) to the data, using the values of $a$ and $b$ determined by fitting the usual GR (red) line to the small magnitude end between $3.0 \leq M < 4.5$. We see that $b = 0.93 \pm .02$, implying that $\tau \sim 2.61$, near the value of $\tau \sim 2.5$ characteristic of mean field and near mean field systems. Likewise, we find that $\sigma = 1.0 \pm 0.37$, which is the value for mean field systems. We carry out a similar analysis for cycles of earthquakes larger than M≥ 6 and show the results in Figure 5. Fitting the data between $3.0 \leq M < 5.0$ we find that $b = 0.90 \pm .01$, thus $\tau \sim 2.67$. In addition, $\sigma = 0.99 \pm 0.38$ as before.

We point out that many of the earthquakes shown on the maps of Figure 3 are located far offshore, where there are no seismometers to record small earthquakes. We would therefore expect that the completeness threshold for many of these offshore

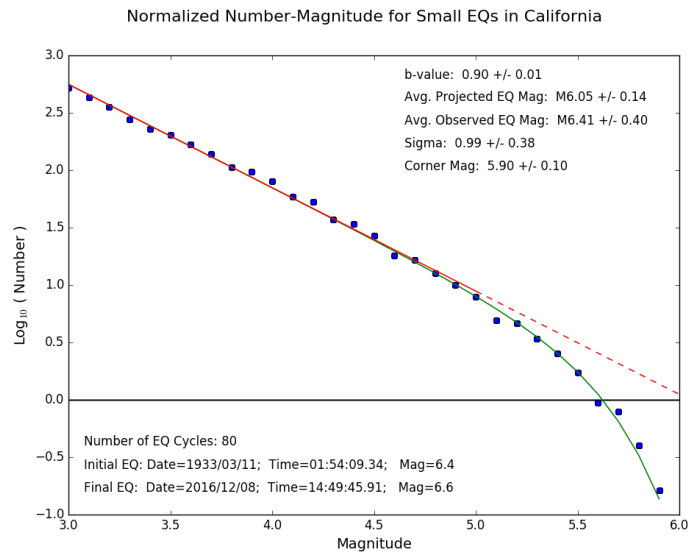

Figure 5. Average GR number-magnitude statistics for small earthquakes $3 \leq M < 5$ in California in the region shown in Figure 3a. Red line shows a fit to the data between magnitudes $3 \leq M \leq 4.5$. Green line is a fit to equation (5), the droplet model

locations would be at magnitudes larger than 3.0. The implication is that the actual *b* value for the scaling line might be closer to $b \sim 1$, and therefore that $\tau \sim 2.5$.

The fact that these scaling exponents, $\tau$ and $\sigma$, have values close to to their mean field values, indicates that seismicity probably has the ergodic property, and thus that time averages should be equal to ensemble or spatial averages. In turn, this means that the fundamental assumption of nowcasting, and of the forecasting method described in Holliday et al. (2014), has a sound basis. Or put more simply, that a deficiency of large earthquakes in the GR statistics, which are not present in Figure 5, must eventually be filled in by the occurrence of large earthquakes.

## Application to Megacities: Nowcasting

*Method.* In earthquake fault systems, a critical problem is the estimation of the state of stress in the earth, its relation to the failure strength of the active faults in a region, and the rate of stress accumulation (Scholz, 1990). Determining the values of these parameters would allow researchers to estimate the proximity to failure of the faults in the region. This would be an answer to the question of "how far along is the region in the earthquake cycle?". However, it is not possible to determine the state of stress to anywhere near the accuracy needed to answer this question. What can be done is to use proxy data, counts of small earthquakes since the last large earthquake in local regions, and use the statistics of regional events to relate the count of small earthquakes to progress through the regional earthquake cycle.

Our goals is to assign a number to a defined small geographic region at a given time, to characterize the current potential for a "large" earthquake to occur in the region. We denote the magnitude of the "large" earthquake as $M_\lambda$. "Large" in this sense means having the potential to cause damage or injuries if it were to occur close by. We compute the potential by using "small" earthquakes, whose magnitude is denoted by $M_\sigma$. The large earthquake magnitude is selected by a procedure described below, whereas the small earthquake magnitude is typically set by the catalog completeness level, but can be also viewed as a free parameter within a certain range.

The Gutenberg-Richter magnitude-frequency relation (Gutenberg and Richter, 1942) can be used to show that the number of small earthquakes having magnitudes larger than $M_\sigma$ but less than magnitude $M_\lambda$ is on average a known value $N$. We can calculate $N$ using the Gutenberg Richter law for the average number $N_{avg}$ of earthquakes greater than $M$:

$$N_{avg} = 10^a 10^{-bM} \qquad (8)$$

where the *b*-value is typically a number near 1 and the value of *a* is set by the level of seismicity in the region (Scholz, 1990).

If we denote by $n_\sigma$ the number of earthquakes having magnitude larger than $M_\sigma$, and by $n_\lambda$ the number of earthquakes having magnitude larger than $M_\lambda$, we find:

$$N = \frac{n_\sigma - n_\lambda}{n_\lambda} \qquad (9)$$

$$N = 10^{b(M_\lambda - M_\sigma)} - 1 \qquad (10)$$

As the number of small earthquakes since the last large earthquake increases with time, the likelihood or potential for another large earthquake increases. The number of small earthquakes is often used as a measure of the passage of *natural time* for the earthquake system (Varotsos et al., 2005; Holliday et al., 2006).

Mathematically, we can express the potential for a large earthquake to occur having magnitude larger than $M_\lambda$ by computing the cumulative distribution function (CDF) of small earthquakes of magnitude larger than $M_\sigma$ but less than $M_\lambda$: $M_\sigma \leq M \leq M_\lambda$. This CDF can computed by considering a large geographic region with a substantial number of earthquakes having magnitude $M_\lambda$ or larger. The number of large earthquake "cycles" is then 1 less than the number of large earthquakes.

To compute the cumulative distribution, we tabulate the number of small earthquakes for each large earthquake cycle from an appropriate database such as the ANSS earthquake catalog (ANSS, 2017), then use this to define the probability density function (PDF) and the cumulative distribution function (CDF) by standard methods (e.g., Bevington and Robinson, 2003).

Once we have the CDF, we can then use the current count of small earthquakes $n(t)$ at time $t$ to compute the current value of the CDF, $P\{n \leq n(t)\}$. Note that $t$ is the (calendar) time of the last large earthquake, and $n(t)$ is the number of small earthquakes since the last large earthquake. This value is then assigned at time $t$ as the *Earthquake Potential Score* (EPS):

$$\text{EPS} \equiv P\{n \leq n(t)\} \qquad (12)$$

We interpret the EPS as the progress of the small area through the earthquake cycle, and thus the potential for the occurrence of the next large earthquake having magnitude larger than $M_\lambda$. By construction, EPS will increase with monotonically with time since the last large earthquake. It will reset to $\text{EPS} = P\{n \leq n(t=0)\} = 0$ immediately after the next large earthquake, and then again begin to increase monotonically until the next large earthquake occurs. We note that the nowcasting method does not, a priori, involve a model, it is only a process of tabulating and interpreting data. However, to the extent that the results have meaning, it is a transparent way of estimating the progress of a region through the seismic cycle of large earthquakes.

As an example of these calculations, Figure 6 shows a map within which is the city of Tokyo Japan, embedded at the center of a 4000 x 4000 km geographic region. A blue circle of radius 200 km surrounds Tokyo on the map (the circle appears as an ellipse due to the map projection). The large region is used to collect the statistics for the nowcast calculation. Small earthquakes within the circle that have occurred since the last M>6.5 large earthquake on 07/30/2011 are used to compute the nowcast. All earthquakes, small and large, are constrained to have depths no larger than 100 km.

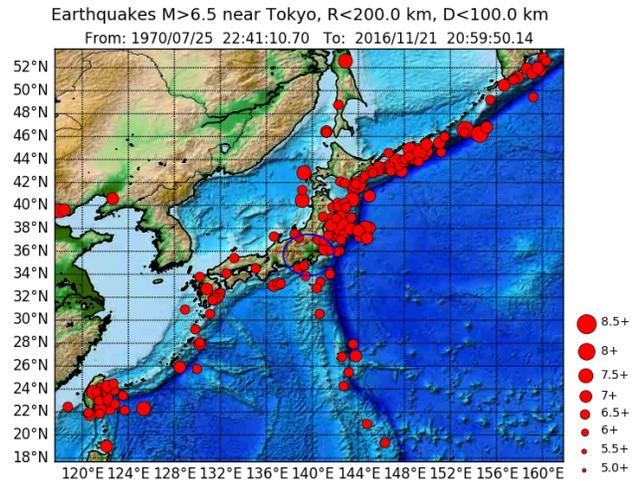

Figure 6. Epicenters of 136 earthquakes $M \geq 6.5$ since 1970 in a region of size 4000 x 4000 km at depths less than 100 km near Tokyo, Japan. Blue ellipse is a circle of radius 200 km surrounding Tokyo.

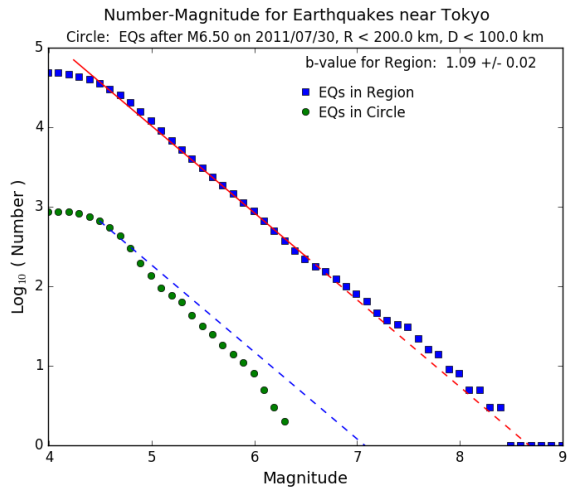

Figure 7. GR number-magnitude statistics for the region in Figure 6. The upper blue square symbols are all earthquakes $M \geq 4$ for the region as a whole since 1970. The lower green circles are all earthquakes $4 \leq M < 6.5$ since the last $M \geq 6.5$ earthquake within the 200 km radius circle around Tokyo, an M6.5 event on 7/30/2011.

The corresponding GR number-magnitude statistics are shown in Figure 7. Note that the completeness magnitude, which has been assumed to be $M \geq 4.0$, is actually closer to M~4.5 for the Tokyo region. Recall that whereas the blue square symbols and red line in Figure 7 represent all earthquakes occurring in the large geographic region, the green circle symbols and blue line represent only the earthquakes occurring within the circular region since the most recent $M \geq 6.5$ earthquake in the circle, which was an M6.5 event on 07/30/2011. Also recall that the blue dashed line has the same slope, $b$=1.09, as the red dashed line by definition.

Figure 8 shows the Earthquake Potential Score for the circular Tokyo region. There have been 917 small earthquakes since the last M6.5 event, corresponding to an EPS score of 92.2%. By this measure, the Tokyo region is near the end of its current earthquake cycle for events $M \geq 6.5$.

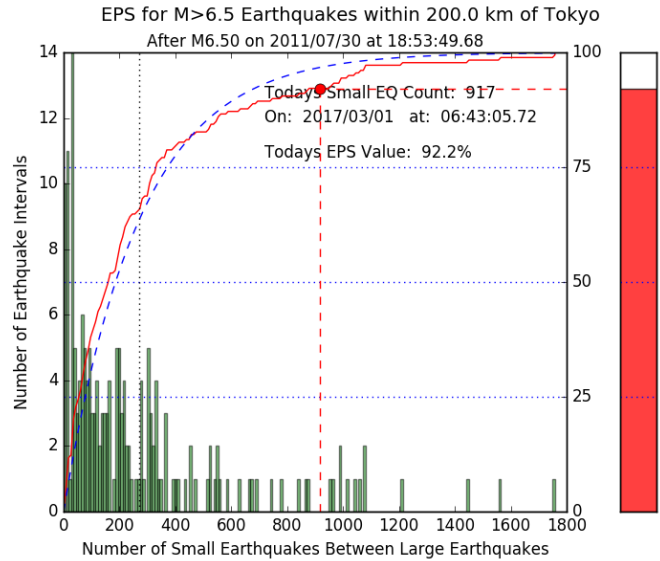

Figure 8. Nowcast for $M \geq 6.5$ earthquakes within the 200 km circle around Tokyo, Japan shown in Figure 7. Green bars are a histogram of the number of small earthquakes arising from the 135 earthquake cycles in the large region. Red curve is the CDF, blue dashed curve is the Poisson CDF with the same mean. Red dot is the current number of small earthquakes since the last large M6.5 event on 7/30/2011 in the blue circle. Red thermometer on the right side is a pictorial representation of the current state of the earthquake cycle.

There were 179 large earthquakes in Figure 6, thus 178 intervals, which are represented by the binned bars in the histogram. The increasing red curve is the CDF $P\{n \leq n(t)\}$ constructed from the 178 large earthquake intervals. The dashed blue curve is the Poisson distribution having the same mean as the CDF $P\{n \leq n(t)\}$. The red dot marks the position of the current small earthquake count. There have been 917 earthquakes in the region having magnitudes $M_\sigma \geq 4.0$ since the last $M_\lambda \geq 6.5$ event on 2011/07/30. The red "thermometer" on the right side is a visual representation of the current cumulative probability (EPS), which is EPS = 92.2%.

***Automated Web Application***. To expedite the process of computing and comparing the current relative risk level of various global cities, we have developed a simple automated web application using Python, HTML, CSS and Javascript. The workflow for this app is:

- Python code is invoked to download the data from ANSS, tabulate the nowcast CDF, build and store the images for epicenter map, nowcast, number-frequency relations, and an XML file that is later read by the web app

- The HTML, CSS and Javasscript is then used to access the XML file, display the images, and display and rank the data in tabular form

A screenshot of this site is shown in Figure 9 for a large region of 4000 x 4000 km around Taipei, Taiwan. In the next section, we discuss results using this app, and in particular we carry out a simple sensitivity analysis of nowcasting parameters.

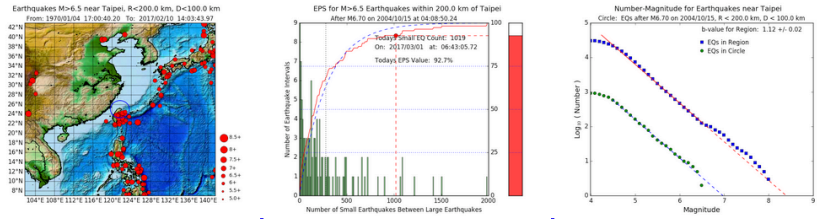

Earthquake Potential Score (EPS) is the percentage of progress through the earthquake cycle. It is possible to get a score over 100%, which would mean the earthquake is overdue, based on historic statistics. The city location is at the center of the 200.0 km radius area.

| Location (Center of Circle) | EPS(%) Score on 2017/04/05 at 13:19:33 PT | Date Last EQ | Mag Last EQ | Natural Time Count on 2017/04/05 at 13:19:33 PT | Mean Count | Std Dev | Number Large EQs in Region |
|---|---|---|---|---|---|---|---|
| Taipei | 92.7 | 2004/10/15 | 6.70 | 1019 | 279 | 353 | 110 |
| Tokyo | 92.2 | 2011/07/30 | 6.50 | 917 | 268 | 322 | 179 |
| Manila | 91.8 | 1999/12/11 | 7.30 | 764 | 265 | 370 | 146 |
| San Francisco | 84.1 | 1989/10/18 | 6.90 | 2079 | 1012 | 1054 | 44 |
| Los Angeles | 82.4 | 1999/10/16 | 7.10 | 1572 | 785 | 729 | 68 |
| Lima | 70.6 | 2007/08/15 | 8.00 | 239 | 173 | 168 | 68 |
| Ankara | 27.0 | 1999/11/12 | 7.20 | 140 | 507 | 670 | 37 |

Figure 9. Screen shot of the web app described in the text. Data in the table represent the 7 megacities, the current nowcast, and statistical data as described in the accompanying text.

The first column of the Table in Figure 9 contains the city at the center of the 200 km circle embedded within a region of size 4000 km x 4000 km. The second column contains the nowcast EPS score computed using the catalog on the date and time listed at the top. The third column contains the date of the most recent $M \geq 6.5$ earthquake within the city circle, and the fourth column contains its magnitude. The fifth column contains the count of small earthquakes since the last large earthquake within the city circle. The sixth column is the mean number of small earthquakes between large earthquakes in the large geographic region, and the seventh column is the standard deviation. The eighth and final column is the number of large earthquakes within the large geographic region, so the number of earthquake cycles is 1 less than the number of large earthquakes..

***Sensitivity Analysis.*** Here we seek to determine the sensitivity of EPS values to changes in the parameters characterizing the statistics, in particular the size of the region from which the basic statistics are drawn. Tables 1-3 show data computed for a large region of size 3000 km x 3000 km, 2000 x 2000 km, and 1500 x 1500 km respectively.

### Table 1
**EPS Statistics for a large region of size 3000 x 3000 km**
**EPS volume for data is 200 km radius circle, depth 100 km**
**Large earthquakes are magnitudes M≥ 6.5**

| Location (Center of Circle) | EPS(%) Score on 2017/04/05 at 10:58:53 PT | Date Last EQ | Mag Last EQ | Natural Time Count on 2017/04/05 at 10:58:53 PT | Mean Count | Std Dev | Number Large EQs in Region |
|---|---|---|---|---|---|---|---|
| Taipei | 95.2 | 2004/10/15 | 6.70 | 1019 | 233 | 355 | 63 |
| Tokyo | 93.4 | 2011/07/30 | 6.50 | 917 | 279 | 322 | 136 |
| Manila | 89.5 | 1999/12/11 | 7.30 | 764 | 279 | 349 | 95 |
| San Francisco | 81.1 | 1989/10/18 | 6.90 | 2079 | 1151 | 1078 | 37 |
| Lima | 78.3 | 2007/08/15 | 8.00 | 239 | 180 | 231 | 46 |
| Los Angeles | 65.6 | 1999/10/16 | 7.10 | 1572 | 1287 | 1073 | 32 |
| Ankara | 29.0 | 1999/11/12 | 7.20 | 140 | 472 | 620 | 31 |

### Table 2
**EPS Statistics for a large region of size 2000 x 2000 km**
**EPS volume for data is 200 km radius circle, depth 100 km**
**Large earthquakes are magnitudes M≥ 6.5**

| Location (Center of Circle) | EPS(%) Score on 2017/04/05 at 10:51:25 PT | Date Last EQ | Mag Last EQ | Natural Time Count on 2017/04/05 at 10:51:25 PT | Mean Count | Std Dev | Number Large EQs in Region |
|---|---|---|---|---|---|---|---|
| Tokyo | 94.6 | 2011/07/30 | 6.50 | 917 | 261 | 308 | 111 |
| Manila | 93.0 | 1999/12/11 | 7.30 | 764 | 246 | 364 | 57 |
| Taipei | 89.7 | 2004/10/15 | 6.70 | 1019 | 265 | 406 | 39 |
| Lima | 72.4 | 2007/08/15 | 8.00 | 239 | 173 | 180 | 29 |
| Los Angeles | 65.5 | 1999/10/16 | 7.10 | 1572 | 1209 | 1011 | 29 |
| San Francisco | 60.9 | 1989/10/18 | 6.90 | 2079 | 1687 | 1251 | 23 |
| Ankara | 23.8 | 1999/11/12 | 7.20 | 140 | 414 | 410 | 21 |

### Table 3
**EPS Statistics for a large region of size 1500 x 1500 km**
**EPS volume for data is 200 km radius circle, depth 100 km**
**Large earthquakes are magnitudes M≥ 6.5**

| Location (Center of Circle) | EPS(%) Score on 2017/04/05 at 11:14:17 PT | Date Last EQ | Mag Last EQ | Natural Time Count on 2017/04/05 at 11:14:17 PT | Mean Count | Std Dev | Number Large EQs in Region |
|---|---|---|---|---|---|---|---|
| Tokyo | 93.1 | 2011/07/30 | 6.50 | 917 | 286 | 328 | 72 |
| Taipei | 88.9 | 2004/10/15 | 6.70 | 1019 | 287 | 355 | 27 |
| Manila | 77.4 | 1999/12/11 | 7.30 | 764 | 334 | 433 | 31 |
| Lima | 68.4 | 2007/08/15 | 8.00 | 239 | 197 | 184 | 19 |
| San Francisco | 57.1 | 1989/10/18 | 6.90 | 2079 | 1685 | 1283 | 21 |
| Los Angeles | 55.0 | 1999/10/16 | 7.10 | 1572 | 1621 | 1409 | 20 |
| Ankara | 33.3 | 1999/11/12 | 7.20 | 140 | 449 | 541 | 12 |

The exact choice of parameters will in practice be determined by the uses to which the EPS values are put. All data are computed for a volume containing each city of radius 200 km, depth 100 km, and for large earthquakes of magnitude $M \geq 6.5$. Thus we compute EPS values for $M \geq 6.5$ earthquake cycles in the following tables (1-4) that are of the same format of Figure 9. We then compile the comparisons in Table 5.

Note that the assumed completeness magnitude for the two US cities, Los Angeles and San Francisco, is M3.0, whereas the completeness magnitude for the other cities was set at M4.0. This difference has an interesting effect as we now discuss. Examining Tables 1-4, it can be seen that the EPS scores for Tokyo, Taipei, Manila, Lima and Ankara are all fairly consistent, but that the EPS scores for San Francisco and Los Angeles are much larger at the largest geographic region sizes (4000 x 4000 km and 3000 x 3000 km) than they are at the smallest regions sizes (2000 x 2000 km and 1500 x 1500 km).

| Table 4 EPS Scores for 7 Cities and Regions of Varying Sizes for M6.5 Earthquakes ||||||
|---|---|---|---|---|---|
| Location | 4000 x 4000 km | 3000 x 3000 km | 2000 x 2000 km | 1500 x 1500 km | Average |
| Completeness Magnitude: M4.0 ||||||
| Ankara | 27.0 % | 29.0 % | 23.8 % | 33.3 % | 28.3 ± 3.4 % |
| Lima | 70.6 % | 78.3 % | 72.4 % | 68.4 % | 72.4 ± 3.7 % |
| Manila | 91.8 % | 89.5 % | 93.0 % | 77.4 % | 87.9 ± 6.2 % |
| Taipei | 92.7 % | 95.2 % | 89.7 % | 88.9 % | 91.6 ± 2.5 % |
| Tokyo | 92.2 % | 93.4 % | 94.6 % | 72.0 % | 88.1 ± 9.3 % |
| Los Angeles | 57.4 % | 50.0 % | 62.1 % | 50.0 % | 54.9 ± 5.2 % |
| San Francisco | 56.8 % | 54.1 % | 39.1 % | 42.9 % | 48.2 ± 7.4 % |
| Completeness Magnitude: M3.0 ||||||
| Los Angeles | 82.4 % | 65.6 % | 65.5 % | 55.0 % | 67.1 ± 9.8 % |
| San Francisco | 84.1 % | 81.1 % | 60.9 % | 57.0 % | 70.8 ± 12.0 % |

A difference like this is what one might expect if, in the larger regions, events of magnitude $M \sim 3$ were not recorded. In fact, these larger regions in Figure 9 and Table 1 are comprised of a large contribution from offshore earthquakes, which are relatively distant from the nearest land-based seismometers, therefore having a higher expected completeness magnitude. The smaller earthquakes would likely not be recorded as easily as when these events occur on land. Table 4 shows EPS summary data computed for a large region of varying sizes with two different completeness magnitudes for Los Angeles and San Francisco. Completeness magnitude $M \sim 4$ is the more consistent choice.

## Discussion and Summary

We have discussed the idea of natural time and applied it to the problem of identifying the current seismic state of a fixed geographic region within its local earthquake cycle, a problem that we refer to as nowcasting. Rather than focus on a specific fault or several faults, as is typical in geophysics and geology, we focus on the defined set of faults encompassed within the geographic area. This method avoids the developing problem now being recognized, that earthquakes on complex fault systems can jump from fault to fault in complex patterns that are essentially unpredictable in advance (Hamling et al., 2017).

Nowcasting is in fact a concept borrowed from economics and finance. Once the current state of the fault system in the defined area is better understood, it should be possible to better characterize the future state of the region and the calculation of forecast probabilities.

In our nowcasting method, we construct the cumulative probability distribution function (CDF) for the number of small earthquakes between large earthquakes during a sequence of earthquake cycles. These cycles occur in a large region around the area of interest, typically a local region around a specific geographic location. An advantage of this method is that it offers a systematic means of ranking locations as to their current exposure to the earthquake hazard. We calculate an Earthquake Potential Score (EPS) that is found from determining the number of small earthquakes since the last large earthquake, then using the CDF found from the regional earthquake cycles. Specifically, if $n(t)$ is the number of small earthquakes since the last large earthquake, the EPS score is defined to be: $\text{EPS} = P\{n \leq n(t)\}$, where $P$ is the CDF of small earthquakes occurring between large earthquakes.

We also related the result from previous studies, that observed earthquake seismicity indicates ergodic dynamics, that ensemble (spatial) averages should be interchangeable with temporal averages. This implies that the seismicity statistics of local regions over long times should be similar to the seismicity statistics over larger regions on much shorter time scales.

Practical applications of these ideas are straightforward. We constructed an automated web application using HTML, Javascript, CSS and Python to implement the nowcasting paradigm. We then showed that rankings of global cities could be constructed based on ANSS catalog data. We carried out a sensitivity analysis to determine the dependence of hazard rankings on large geographic region size and completeness magnitude. We found that a completeness magnitude of M4.0 gave the most consistent results for the US cities of Los Angeles and San Francisco, as well as for the global cities of Ankara, Lima, Manila, Taipei and Tokyo. For these calculations, we defined the "local" region as a region of radius 200 km around the city, and a maximum earthquake depth of 100 km for hazard from large earthquakes having magnitudes $M \geq 6.5$.

We conclude that our method offers a rapid and easily reproducible method to estimate the current level of progress through the earthquake cycle for any local region subject to seismic hazard worldwide.

**Acknowledgements**.  Research by JBR, ML, and AG were supported under NASA grant NNX12AM22G to the University of California, Davis.